\documentstyle[epsf]{article}
\newskip\humongous \humongous=0pt plus 1000pt minus 1000pt

\newif\ifdtup

\catcode`\@=11
 
\@addtoreset{equation}{section}
\def\theequation{\arabic{equation}}
  
\def\@normalsize{\@setsize\normalsize{15pt}\xiipt\@xiipt
\abovedisplayskip 14pt plus3pt minus3pt%
\belowdisplayskip \abovedisplayskip
\abovedisplayshortskip \z@ plus3pt%
\belowdisplayshortskip 7pt plus3.5pt minus0pt}
 
\def\small{\@setsize\small{13.6pt}\xipt\@xipt
\abovedisplayskip 13pt plus3pt minus3pt%
\belowdisplayskip \abovedisplayskip
\abovedisplayshortskip \z@ plus3pt%
\belowdisplayshortskip 7pt plus3.5pt minus0pt
\def\@listi{\parsep 4.5pt plus 2pt minus 1pt
     \itemsep \parsep
     \topsep 9pt plus 3pt minus 3pt}}
 
\relax

\catcode`@=12
 
\catcode`\@=11
 
\def\section{\@startsection{section}{1}{\z@}{3.5ex plus 1ex minus
   .2ex}{2.3ex plus .2ex}{\large\bf}}
 
\def\thesection{\arabic{section}.}

\def\appendix{\setcounter{section}{0}
 \def\thesection{Appendix \Alph{section}:}
 \def\theequation{\Alph{section}.\arabic{equation}}}
 
\begin{document}
\begin{titlepage}
\begin{center}
{\LARGE
Mass, Confinement and CP Invariance 
in the Seiberg-Witten Model}
\end{center}

\vspace{1em}
\begin{center}
{\large
Massimo Di Pierro$^{1}$ and  Kenichi Konishi$^{2}$}
\end{center}

\vspace{1em}
\begin{center}
{\large {\it $^{1}$  Dipartimento di Fisica -- Universit\`a di Pisa\\
  Istituto Nazionale di Fisica Nucleare -- sez. di Pisa\\  
 Piazza Torricelli, 2 -- 56100 Pisa (Italy)\\
E-mail:  df144023@ipifidpt.difi.unipi.it   \\

$^{2}$  Dipartimento di Fisica -- Universit\`a di Genova\\
     Istituto Nazionale di Fisica Nucleare -- sez. di Genova\\
     Via Dodecaneso, 33 -- 16146 Genova (Italy)\\
     E-mail:  KONISHI@GE.INFN.IT\\} }
\end{center}

\vspace{7em}
{\bf ABSTRACT:}
{\large Several physics aspects of the Seiberg-Witten  solution of $N=2$ supersymmetric
Yang-Mills theory with $SU(2)$ gauge group, supplemented with a small mass term
for the "matter" fields which leads to an $N=1$ theory with confinement,  are
discussed. The light spectrum of the theory is understood on the basis of current
algebra relations and  the CP invariance of the massless
and massive theories are studied.  We  find  that in the massive (confining) theory
the low  energy physics   has an exact CP symmetry, while in a generic vacuum in
the massless theory CP invariance is broken spontaneously. }

\vspace{2em}
\begin{flushleft} 
{\large GEF-Th-6/1996~~~~~~~~~~~~~~~~~~~~~~~~~~~~~~~~~~~~~~~~~~~~~~~~~~~~~~ May 1996}
\end{flushleft} \end{titlepage}

\def\MQ{meccanica quantistica~}
\def\MC{meccanica classica~}
\def\beq{\begin{equation}}
\def\eeq{\end{equation}}
\def\non{\nonumber}
\def\De{\Delta}
\def\bea{\begin{eqnarray}}
\def\eea{\end{eqnarray}}
\def\bra{\langle}
\def\ket{\rangle}
\newcommand{\defi}{\stackrel{\rm def}{=}}
\def\de{\partial}
\def\si{\sigma}
\def\sb{{\bar \sigma}}
\def\rn{{\bf R}^n}
\def\r4{{\bf R}^4}
\def\s4{{\bf S}^4}
\def\Tr{\hbox{\rm Tr}}
\def\ker{\hbox{\rm ker}}
\def\dim{\hbox{\rm dim}}
\def\sup{\hbox{\rm sup}}
\def\inf{\hbox{\rm inf}}
\def\re{\hbox{\rm Re}}
\def\im{\hbox{\rm Im}}
\def\infi{\infty}
\def\nrm{\parallel}
\def\nrmi{\parallel_\infty}
\def\teo{\noindent{\bf Theorem}\ }
\def\bra{\langle}
\def\ket{\rangle}
\def\all{\hbox{\rm all}}
\def\co{$^,$}
\def\daa{$^-$}
\def\dirac{{\cal D}}
\def\dplus{{\cal D_{+}}}
\def\dminus{{\cal D_{-}}}
\def\om{\Omega}
\def\me{m_e}
\def\mp{m_p}
\def\e2{e^2}
\def\r2{r^2}
\def\cm{{\rm cm}}
\def\erg{{\rm erg}}
\def\se{{\rm sec}}
\def\gram{{\rm gr}}
\def\mol{{\rm mol}}
\def\cost{\hbox {\rm cost.}}
\def\o{\over}

 In a   celebrated work\cite{SW} Seiberg and Witten  exploited the (generalized)
electromagnetic duality, $N=2$ supersymmetry and  holomorphic 
property of effective actions, 
to solve   exactly   a strongly interacting
non Abelian theory in four dimensions, i.e.,  to compute the vacuum degeneracies,
and in each vacuum, to determine  the exact spectrum  and   interactions among light 
particles. 

An especially interesting observation of Ref \cite{SW}, made in the  pure $N=2$
supersymmetric Yang-Mills theory with $SU(2)$ gauge group, 
 is that upon turning 
on the mass term\footnote{We follow the notation of Ref\cite{SW}.}
 $\int d^2\theta \,Tr\, m\Phi^2, $   $m
\ll \Lambda, $   the light  magnetic monopole field
condenses,  providing thus the first explicit  realization of the confinement
mechanism envisaged by  't Hooft.\cite{TH}  In this note we wish to make some 
further
  observations on this model.

There are in fact several related questions which to our knowledge have not
yet been  discussed fully. 
Why do the vacua in the massive (confining) theory 
correspond precisely to those points  of the quantum moduli space (QMS) of the $N=2
$ theory where the magnetic monopole becomes massless?  Usually,  one expects that a
dynamically generated mass in a non Abelian theory with scale  $ \Lambda$   is of the
order of  $\Lambda$, while in this  model   the mass gap is of the
order of $m^{1/2} \Lambda^{1/2}/ \log^{1/2}(\Lambda/m)  $ (see below).  Why is that
so?   Is the CP invariance spontaneously broken at low energies?   Does   the
oblique confinement\cite{TH} take place?  Do dyons condense? What is the relation
between the Seiberg-Witten effective action and the more speculative  (but
supposedly exact) effective superpotential  constructed for the $N=1$ theory with
the "integrating in" procedure?\cite{Sei} 
 Finally,  has all this got anything to do with what happens in an  $N=0$ theory
of interest such as  QCD?

\smallskip

A first general remark is that  one is here dealing with a system with  large
vacuum degeneracy, intact after full quantum corrections are taken into account.
This  vacuum degeneracy  is (almost) eliminated    by the
mass perturbation,  $\int d^2 \, \theta \,Tr \,m\Phi^2,$  leaving only  a double
degeneracy corresponding to the order parameter $ u = \bra \Tr \, \phi^2 \ket =
\pm \Lambda^2.$ 

 As is usual in the  standard degenerate perturbation theory in quantum
mechanics, to lowest order  the only effect of the perturbation is to  fix the
vacuum to the "right",    but  unperturbed,   one, with $N=2$ (hence $SU_R(2)$)
symmetric properties.  This means that  the  $SU_R(2)$ current of the original
theory, \beq   J_{\mu}^a = \Tr \,{\bar \Psi} {\bar \sigma_{\mu}} \tau^a \Psi
\label{currenthigh}\eeq
 where $\Psi = \pmatrix{\lambda \cr \psi},$   is
well approximated  in the low energy effective theory by the Noether currents
(of the low energy theory)\footnote{The effective low energy action 
 is \cite{SW}
${1\over 4\pi} Im \, [ \int d^4\theta \, {(\de F(A_D)/\de A_D)} {\bar
A_D} +  \int d^2 \theta \, {(\de^2 F(A_D)/ \de A_D^2)} W_D W_D/2 \,],  $
   with the addition of the standard terms for the magnetic monopole
sector, $  M^{\dagger}e^{V_D} M + {\tilde M}^{\dagger}e^{-V_D} {\tilde M}|_{D}+ 
\sqrt{2} A_D {\tilde M} M|_F +h.c.$  The mass term is geiven by
$\int d^2 \theta   \, m \,\Phi^2 + h.c.= \int d^2 \theta   \, m \,U(A_D) +
h.c.$  }

  \beq J_{\mu}^a = {\bar \Psi_D} \,{\bar
\sigma_{\mu}} \tau^a \Psi_D + i (D_{\mu} S^{\dagger})  \tau^a  S -
i  S^{\dagger}  \tau^a  D_{\mu} S,
\label{currentlow}\eeq  
where $\Psi_D = \pmatrix{\lambda_D \cr \psi_D}$  is the dual of the (color diagonal 
part of)  $\Psi$,  and $S\equiv \pmatrix { M \cr {\tilde M}^{\dagger}}$.
 In arriving at  Eq.(\ref{currentlow}) we
identified  the high energy and low energy $SU_R(2)$ groups such that  the doublet
$\pmatrix{\lambda_D \cr \psi_D}$ transforms as $ \pmatrix{\lambda \cr \psi}$.

Let us consider the current-current correlation function 
\beq i \int d^4 x \, e^{-iqx} \bra 0| T\{ J_{\mu}^3(x) J_{\nu}^3(0)\}|0\ket
= (q_{\mu}q_{\nu} - q^2 g_{\mu\nu} ) \Pi(q^2), \label{correl}\eeq
and an associated $R$-like ratio, 
\beq   R= Disc_{q^2} \Pi(q^2), \eeq
appropriately normalized (by introducing  hypothetic "leptons").  At high energies $R$
just counts the number of $\lambda$ and $\psi$  "quarks" (asymptotic freedom), apart
from calculable logarithmic  corrections as well as an infinite number of
not-so-easily-calculable power corrections (involving  gluonic and higher
condensates): 
\beq R_{q^2 \gg \Lambda^2} \simeq  6 +  O(\alpha(q^2),
q^4/\Lambda^4).\eeq
 At low energies, the same quantity is simply given by the weakly interacting dual
fields and  magnetic molopoles,
\beq R_{m\Lambda  \ll q^2 \ll \Lambda^2} \simeq  3, \eeq
where two out of three comes from the contributions from $\lambda_D$ and $\psi_D$ (one
each) and one  from $M$ and ${\tilde M}$ (one half each).  This amounts to
an exact resummation of an infinite number of power and  logarithmic corrections.

To next order the effect of the explicit  $SU_R(2)$  breaking,
\beq \de^{\mu} J_{\mu}^{-}= \de^{\mu}  \Tr \, {\bar \psi } {\bar
\sigma}_{\mu} \lambda = i\, m \, \Tr \, { \lambda }{\psi}; \eeq
\beq \de^{\mu} J_{\mu}^{3}= {1\over 2} \de^{\mu}  \Tr \, ({\bar \lambda } {\bar
\sigma_{\mu}}\lambda -  {\bar \psi  } {\bar
\sigma_{\mu}}\psi) = {i\over 2}( \Tr \, m \psi^2 - h.c. )\eeq
 must be
taken into account.\footnote{Although we are discussing here the first order mass
corrections  in a degenerate perturbation theory,  many  crucial relations
are  exact due to the  nonrenormalization theorem (the form of the
superpotential, etc). Most results  below, in particular the
exact CP invariance of the low energy theory,  should survive higher order 
corrections which affect only the  D type  terms.}  
In particular,     the  anomaly of Ref \cite{KK}
\beq   m\,u = m\, \bra  \Tr \,\phi^2 \ket  = 2 {g^2\over 32\pi}\bra \Tr \,
\lambda(x) \lambda(x)\ket, \label{konishi}\eeq  hence 
\beq {(g^2/ 32\pi)}\bra \Tr \,\lambda(x) \lambda(x)\ket = \pm m \Lambda^2/2, \eeq 
implies that  the  $SU_R(2)$ symmetry is  spontaneously broken also (which may be 
called {\it induced} breaking).\footnote{This is similar to what happens to  the
isospin invariance in QCD: although the main dynamical effect (quark condensation) is
isospin invariant,  the addition of unequal up and down quark masses causes  the
respective   condensates to take a little  different values, thus breaking
spontaneouly the isopspin, very slightly.} 
As  $\lambda^2$ condensate is a triplet it breaks $SU_R(2)$ symmetry completely,
implying {\it three} light pseudo Nambu-Goldstone particles.  Also,  the 
supersymmetric Ward-Takahashi  like identity 
\beq    \bra \Tr \, \psi^2 \ket =2 \,m^{*}
\bra  \Tr \,\phi^{*} \phi  \ket, \label{psicond}\eeq
 and the fact that $\phi$ is in the
adjoint representation hence probably $ \bra  \Tr \,\phi^{*} \phi  \ket \sim \cost
\bra  \Tr \,\phi^2  \ket \simeq \Lambda^2,$ suggests that $  \bra \Tr \,\psi^2 \ket
\sim O(m\Lambda^2 )$ also. 

The associated "pion decay constants" can be easily read off from the expression
of the low energy currents Eq.(\ref{currentlow}). Let us recall that  upon  turning on
the mass term    the vacuum is found to be fixed at
$A_D=0$, $u=\Lambda^2$    as noted by
Seiberg and Witten.\cite{SW} The superpotential $ \sqrt2 A_D {\tilde M} M + m
\,U(A_D),$ minimized  with respect to $A_D$, ${\tilde M}$ and   $M$, indeed yields 
$A_D=0$ and the magnetic monopole condensation,
\beq   \bra M \ket =  \bra {\tilde
M} \ket= (-{m \,U^{'}(0)/\sqrt2})^{1/2}.  \label{magncond}\eeq
Expanding the magnetic monopole fields around its vacuum expectation value 
 \beq  M= \bra M \ket + M^{'},
\quad  {\tilde M}= \bra M \ket +{\tilde M}^{'},   \label{shift}\eeq 
  one 
finds that 
\beq   F_{\pi}^{-} \sim  F_{\pi}^{3} \sim  \bra M \ket =O(m^{1/2}
\Lambda^{1/2}). \label{fpi}\eeq 
Also, it is easy to see that the three pseudo Nambu
Goldstone bosons are (to lowest order) the real and imaginary parts of $M -
{\tilde M}$ and the imaginary part of  $  \bra M^{\dagger}\ket (M + {\tilde
M})$, apart from normalization.   
(Actually,  a linear combination of the real and imaginary parts of 
$M - {\tilde M}$ becomes the longitudinal part of the dual vector boson by the Higgs
mechanism.)

As for the masses of the light particles,  they  can be  studied most easily 
from the fermion bilinear terms arising from the Yukawa interaction terms
upon shifting the magnetic monopole fields  (Eq.(\ref{shift})).  It reads
\bea   L_{Y} &=&  \sqrt2 [ - \bra M^{\dagger} \ket \lambda_D (\psi_M -\psi_{\tilde
M}) + \bra M \ket \psi_D ( \psi_{\tilde M}  + \psi_M) ] 
+{m \over 2} U^{''}\!(0) \psi_D  \psi_D ] + \non \\
 &+& \sqrt2 [ - M^{\prime \dagger } \lambda_D \psi_M +
 {\tilde M}^{\prime  \dagger} \lambda_D \psi_{\tilde M} - 
 (A_D \psi_M \psi_{\tilde M} + \psi_D \psi_{\tilde M} M^{'} +  \psi_D \psi_M {\tilde
M}^{'})]\non \\ 
 &+&
 {m \over 2} (U^{''}\!(A_D)- U^{''}\!(0)) \psi_D  \psi_D  
\,\,+\,\, h.c.\label{Yukawap}\eea 

 A subtle point in reading off the masses from Eq.(\ref{Yukawap}) is
that since the fields   $\lambda_D$   and $\psi_D$ do not have the canonical kinematic
terms, they must be re-normalized    by $\lambda_D \to  g_D \lambda_D$; 
 $\psi_D \to  g_D \psi_D$.   (In the
formula  (\ref{currentlow}) for  the low energy currents  such a rescaling has
already been done.)   In doing so the argument of the dual coupling constant $g_D$ should be
taken as $(m/\Lambda)^{1/2} $ and not $a_D=0$, since in the massive theory there is an
infrared cutoff.  (Such a replacement should automatically take place  if the 
perturbation in $m$  is pushed to higher order.  An analogous phenomenon is known
in the old chiral perturbation theory due to the small pion mass.)   This explains 
the $\log (\Lambda/m)$ dependence of the masses below. 

    From Eq.(\ref{Yukawap}) one  sees that the fields  $\lambda_D$ and $
\psi_2\equiv ( \psi_M - \psi_{\tilde M})/\sqrt2 $ form a Dirac type massive
fermion with mass  \beq    m_1  = |2 g_D \bra M^{\dagger}  \ket |=  2 |g_D
\cdot (\sqrt2 i m \Lambda)^{1/2}| = {4\cdot2^{3/4}\, \pi | m  
\Lambda|^{1/2}\over \log^{1/2}|\Lambda/m|}, \label{mass1} \eeq while in the
subspace of two Weyl fermions $\psi_D$ and $\psi_1\equiv ( \psi_M +
\psi_{\tilde M})/\sqrt2 $ the mass matrix reads  \beq  \pmatrix { {m } g_D^2
U^{''}\!(0)/2 &  g_D \bra M  \ket \cr
  g_D \bra M  \ket & 0 }, \eeq
where $U^{''}\!(0) =-1/2.$
 The phases of these matrix elements can be chosen all  real and
positive by an appropriate phase rotation of the  $\psi_D$ and $\psi_1$ fields
(more about these phases below).   A real symmetric matrix can be diagonalized by a
real orthogonal matrix, leading to mass eigenvalues \beq   m_{2,3} \simeq  \gamma \,|m
\Lambda|^{1/2} \pm {1\over 2} \delta \,|m|,   \label{mass23} \eeq where
\beq   \delta = {g_D^2\over 4};\quad \gamma= 2^{1/4} g_D; \quad 
g_D= {2\sqrt2 \pi \over \log^{1/2}|\Lambda/m|}.\eeq

 The  result,  Eq.(\ref{mass1}), Eq.(\ref{mass23}),  that all the light
particles  have mass,
\newline
 $m_{\pi} \sim O(m^{1/2}\Lambda^{1/2}/ \log^{1/2}(\Lambda/m)),
$ as well as the  result for $F_{\pi}$,  Eq.(\ref{fpi}),   can  be understood in
the light of  the Dashen's formula\cite{RD} 
 \beq   |F_{\pi}|^2 m_{\pi}^2 \simeq  m \bra \Tr \,\psi^2 \ket + h.c.  
\label{Dashen} \eeq
 which is  obtained by saturating the current-current
correlation functions  such as Eq.(\ref{correl}) (at $q^2 \le m \Lambda$) by lowest
lying particles.  Eq.(\ref{Dashen})  is indeed satisfied if on the right hand side 
   $\bra \Tr \,\psi^2 \ket \sim   m\,\Lambda^2 /\log (\Lambda/m)$   which is
quite reasonable in view of Eq.(\ref{psicond}).\footnote{ The generalized Dashen's
formula by Veneziano,\cite{Ven}  with $\sim m  \bra \Tr \,\phi^2 \ket$  on the right
hand side,     here neither   confirms nor contradicts these results.  Due to the
fact that the scalar $\phi$  is neutral with respect to $SU_R(2)$, it simply states
that certain operator has no amplitudes to produce these pseudo Goldstone bosons
from the vacuum.}   Also, the (order-of-magnitude) equality between $F_{\pi}^{-}$
and $F_{\pi}^{3}$  and   between $m_{\pi}^{-}$ and $m_{\pi}^{0}$ is what is
expected:  the same operator $ m \Tr \,\psi^2 + h.c.$ appears on the right hand side
of the Dashen's formula for the correlation functions, $ T\{ J_{\mu}^{+}
\,J_{\nu}^{-}\}$ and $ T\{ J_{\mu}^{3} \,J_{\nu}^{3}\}$.

 The  $N=1$ theory  (with $m \ll
\Lambda$) thus  contains in its spectrum   three light pseudo Nambu-Goldstone bosons
whose masses  vanish  in the limit $m \to 0$  as $\sim m^{1/2}$, and their
($N=1$) superpartners degenerate in mass.  Since low energy chiral superfields    $A_D$
and $W_D$ together contain only  two scalars, we know  {\it a priori}  that there must
be   some  massless scalars  in the low energy effective theory other than these
particles - the situation is saved by the light monopoles (which plays here the role
of pseudo Nambu-Goldstone bosons)!  In other words,  the two  $N=1$ vacua could not be
anywhere else in the QMS of $N=2$ theory,  except at the two singular points $u=\pm
\Lambda^2$.   Other points e.g., with $ |u|=|\Lambda^2|, $  correspond  to theories
with only two massless scalars (see Fig. 1 below)  hence  are not near any of the 
$N=1$ vacua.

     Exactly the same physics arises if one starts with 
an effective action with the $(\pm 1, \pm 1)$  "dyons",  ${\tilde N}, N$,  coupled 
to  $A_D + A$ fields: one finds that the introduction of a small  mass term fixes 
the vacuum to be 
 at  $A_D + A=0$  ($u=e^{+i\pi} \Lambda^2,\,\, \theta_{eff}=-2\pi $).  Because of the
Witten's effect ($q$ is the observed electric charge, $n_e, n_m$ are the integer
quantum numbers labelling the particles)
 \beq     q=  n_e + ({\theta_{eff} /
2\pi}) n_m ,  \label{elcharge}\eeq 
 the $(1,1)$ particle (which would be massive at $u= + \Lambda^2$ but massless at 
$u=e^{+i\pi} \Lambda^2$) 
actually becomes a pure magnetic monopole there and condenses, leading to confinement
of the  color electric charges.   After a redefinition of the low energy fields
by $A_D + A   \to A_D $  (which is an allowed $SL(2, Z)$ transformation) and a 
relabelling of the quantum number   $  n_e \to n_e - n_m$,  the theory in this vacuum
becomes even  formally identical to the case $u=\Lambda^2$   studied above.  
Thus the "dyon condensation" does not occur in this model, contrary to such a 
statement sometimes  made  in the literature.  We may therefore  limit ourselves here
and below to the  $u=\Lambda^2$ vacuum without losing generality.

In view of the light spectrum found here let us compare the massive Seiberg-Witten
effective low energy theory  to  a more na\"{\i}ve  guess for the structure of the 
effective action, constructed   by introducing just two chiral superfields $U=\Tr
\,\Phi^2$ and $S\equiv {g^2\over 32 \pi} \Tr \,WW$ and by using the so-called
"integrating-in" procedure, which leads to the superpotential\cite{Sei}
 \beq   S \log {U^2 \over \Lambda^4} + m U. \label{oldguess} \eeq 
 Eq.(\ref{oldguess}) is such that  the anomalous transformation of the action, $ 4
 \alpha (g^2 / 32 \pi) F_{\mu \nu}{\tilde F}^{\mu \nu}$  (under $\Phi \to
e^{i \alpha } \Phi$),   
 is formally
reproduced.  Such an "effective action" yields upon minimization of the scalar
potential, also the correct results, $  u = \bra \Tr \,\Phi^2  \ket = \pm
\Lambda^2$ and the anomaly Eq.(\ref{konishi}).  Nonetheless, it is an incorrect low
energy action in this theory, as it does not contain light particles with mass $\sim
m^{1/2}\Lambda^{1/2}$.  The reason for such a failure seems to lie in the fact that
the global $SU_R(2)$ symmetry (and its small induced breaking) in the limit of
$m/\Lambda \to 0$ has not  been taken into account properly. 

It is  of particular interest to study  the way CP symmetry  is realized  
in the massless ($N=2$)  and in the massive $(N=1)$ theories, and the relation
thereof.  Although one expects no dependence on the bare $\theta$ parameter 
even in the presence of the $\Phi$ mass term since in
the original theory there is a massless charged fermion ($\lambda$),
\footnote{We thank S. Hsu for pointing out an error in this regards which appeared in
the original manuscript.}
  CP symmetry is realized  at low energies
in the Coulomb phase (massless case)  and in the confining phase 
(massive case) in different ways.   In
the massless theory, the $\theta $ indepence is   assured  by the property of
the exact Seiberg-Witten solution,
\beq a= {\sqrt2 \Lambda \over \pi}\int_{-1}^{1} dx\,
{\sqrt{x-u/\Lambda^2}\over \sqrt{x^2-1}};\quad 
 a_D= {\sqrt2 \Lambda \over \pi}\int_{1}^{u/\Lambda^2} dx\,
{\sqrt{x-u/\Lambda^2}\over \sqrt{x^2-1}},\label{swres}\eeq
in the following sense. 
A shift of $\theta$ by $\Delta \theta$ causes the change in $ \Lambda$ 
as  
\beq  \Lambda \to  e^{i \Delta \theta/4} \Lambda. \label{simulrot1}\eeq
 $\Lambda$ depends on $\theta$ this way since it depends on the bare coupling
constant and bare $\theta$ parameter through the complex "coupling constant"
$\tau= \theta/2\pi + 4\pi i/g^2.$\cite{Shifman} 
But if we now move to a different vacuum by
\beq  u  \to e^{i \Delta \theta/2} u \label{simulrot2}\eeq
 the net change is 
\beq    a \to e^{i \Delta \theta/4} a;\quad a_D \to e^{i \Delta \theta/4} a_D:
\label{rotaead}\eeq
a common phase rotation of $a$ and $a_D$.

Thus all physical properties of an appropriately shifted vacuum $u$ with a new 
value of $\theta$  are the same as in the original theory.
   In particular, the full 
spectrum \cite{SW}, 
 \beq  M= \sqrt2 | n_m a_D + n_e a|.  \label{mass}\eeq 
 as well as the low energy coupling
constant and  vacuum parameter are seen to be unchanged, since
\beq   \tau_{eff} = {\theta_{eff} \o 2 \pi} + { 4 \pi i \o
g_{eff}^2  } = {d a_D \o da}. \eeq 
     In other  words,
the ensemble of theories  represented by the points of QMS,
 taken together,  is
invariant under $\theta \to \theta + \Delta \theta.$ 

(The above argument may be inverted:  the invariance of the theory under 
Eq.(\ref{simulrot1}) and  Eq.(\ref{simulrot2}) can be
interpreted
   as an
indication  that  the   anomalous chiral $U(1)$ transformation property of the
theory is indeed correctly incorporated in the low energy action.)

On the other hand, at fixed generic  $u$, 
physics  depends on $\theta$ (as well
as on the bare coupling constant $g$) non trivially: this is so  because
 the CP invariance is spontaneously broken by $u=\bra \Tr \,\Phi^2 \ket \ne 0$.
Equivalently,  at fixed given bare parameters there are theories with
different value of $u$ and  with inequivalent physics.   For instance,
the spectrum of some stable particles depends on $u$ (hence on the effective low
energy coupling constant $g_{eff}$ and the theta  parameter $\theta_{eff}$), if $u$
is smoothly  changed  along
a semicircular path 
 $u=    e^{i  \alpha }\Lambda^2, \,\, \alpha = 0 \to \pi,  $
as is illustrated  in Fig. 1.\footnote {It is important that such a path is taken
outside the curve on which the ratio $a_D/a$ is real and on which  the spectrum
changes discontinuously, some stable particle becoming unstable, etc.  According to
Ref \cite{Ferrari} it is a near ellipse included entirely in the unit circle
$|u|=|\Lambda^2|$ except at the points $u=\pm \Lambda^2$.}
 Note  in particular  the periodicity of the spectrum in  $\theta_{eff}$ with
periodicity $2\pi$,  in spite of a nontrivial spectral flow. 
Also, the  theories at $u=\pm \Lambda^2$ 
are  characterized  by the 
fact that   $\theta_{eff}=0$  (or $\pm 2\pi$ which is the same as  $0$
because of the periodicity).

 \smallskip

Consider now  the massive theory.  
 Upon  turning on the mass term (even infinitesimal)   the vacuum is
 fixed at $A_D=0$, $u=\Lambda^2$ in which  the effectivelow energy 
 vacuum parameter $\theta_{eff}$ takes the value  zero.  Nonetheless, since $u \ne 0$
(as in other vacua in the Coulomb phase) and also $\bra M \ket \ne 0$,   one
might wonder whether    CP invariance is broken spontaneously at low energies.
In fact this does not occur.

To see that the low energy theory has  indeed an exact CP invariance,  let us go
back to  the Yukawa Lagrangian  Eq.(\ref{Yukawap}).
 By using  Eq.(\ref{magncond}) and the fact that  $U^{'}(0)=-2i
\Lambda$;    $\Lambda = e^{i \theta/4} |\Lambda|$, and $U^{''}(0)= -1/2$
one sees that the only nontrivial phases appear in  $m$  and    $\bra M \ket$, 
 \beq    m= e^{i \alpha} |m|,  \quad \bra M \ket= e^{i \beta}|\bra M \ket|;\qquad 
  \alpha \equiv \arg m,\quad  \beta \equiv {\pi/4} +{\arg m/2} +
{\theta/8},\eeq 
as well as through the expansion of $ U^{''}\!(A_D)- U^{''}\!(0) $
in powers of  $A_D/\Lambda$.   From the exact Seiberg-Witten formula for $a_D(u)$
(Eq.(\ref{swres})) one finds by a  change of the integration variable
 that 
\beq {a_D\o \Lambda} ={i \o \pi}\sum_{n=0}^{\infty}{(-)^n \Gamma^2(n+1/2)\,
(u/\Lambda^2-1)^{n+1} \o  2^{n+1} (n+1)! n!} 
\eeq
with an overall factor $i$ on the right hand side. By inverting the series
one gets 
\beq   U(A_D)=\Lambda^2 (1  + f({A_D/i\Lambda}) ); 
\quad   U^{''}\!(A_D) = -
f^{''}\!({A_D/i\Lambda}) \label{Uprime}\eeq 
 where  $ f(x)= 2x + {(1/4)}x^2 -{(1/
32)} x^3 +\ldots$  and $f^{''}(x)= 1/2 - (3/16)x + \ldots $   are real functions of
(possibly complex) variable $x$. 

  Now  first make the phase rotations
\beq  \psi_D \to e^{-i \alpha/2} \psi_D;\quad \psi_M \to e^{ -i(\beta -
\alpha/2)} \psi_M; \quad \lambda_D \to e^{i(2 \beta - \alpha/2)} \lambda_D ; \quad
M^{'}  \to e^{i \beta}  M^{'}. \label{phaserot}\eeq
($\psi_{\tilde M} $  and ${\tilde M}^{'}$  transforms respectively as $\psi_M$ and 
$M^{'}$). 
 These transformations eliminate   phases from all  masse terms  as well as from the
Yukawa terms  involving  $M^{'}$'s.   On the other hand the Yukawa term
 $A_D \psi_M \psi_{\tilde M} $ 
acquires a phase factor $\exp{-2i(\beta - \alpha/2)} =\exp{(-i\pi/2 - i
\theta/4)}$.   The final rotation \beq A_D \to  e^{i \pi/2
+i \theta/4} A_D \eeq
 however  eliminates this  phase from the Yukawa term
and simultaneously transforms the argument of $f^{''}\!(x) $ as 
${A_D/i\Lambda} \to {A_D/|\Lambda|}$, making   the 
expansion coefficients of $ U^{''}(A_D) -  U^{''}(0)$  in $A_D$  all real.

Thus  the low energy effective theory  is indeed independent of the bare 
 $\theta $ parameter.  More important, 
the spontaneous breakdown of CP invariance   \`a la T.D. Lee\cite{Lee} does not 
occur, all masses and Yukawa  interaction coefficients being real. The low energy
vacuum parameter $\theta_{eff}$ is zero.   This completes the proof of CP invariance
in the low energy  theory.\footnote {As a further check, note  that both sides of
the Dashen's formula Eq.(\ref{Dashen})  is indeed independent of the phase
$\theta$:  $F_{\pi}$ enters  as the absolute value squared, $m_{\pi}$ is real,
and on the right hand side, the condensate $m\bra \psi^2 \ket $ is real as can be 
seen from the supersymmetric Ward-Takahashi identity, Eq.(\ref{psicond}). } 
 Since the low energy
 physics does not depend on  the phase $\theta$,   no oblique confinement 
speculated by t'Hooft (condensation of  (1,0) - (1,1) dyon pairs with opposite
electric charges)\cite{TH}  takes place  in this model.

One might be  tempted to conclude   that, by using a similar argument as above,
"spontaneous CP violation" does   not occur {\it a fortiori} in the $m=0$ theory
either,  since 
 $L_Y$ is much simpler in this case (no explicit  mass term, no  magnetic monopole
condensation). This is not so.   First of all, at a generic vacuum $u \ne \pm
\Lambda^2$   there is  nonzero $\theta_{eff} F_{\mu \nu}{\tilde F}^{\mu \nu}$ term
which breaks CP (this is also a spontaneous breaking since the nonzero  $\theta_{eff}$
is due to the vacuum expectation value of $\re dA_D/dA$).   It is true that one can 
transform away $\theta_{eff}$  by an $SL(2, R)$ transformation of the scalar $A_D \to
A_D + (\theta_{eff}/2\pi) A; $   $A \to A$, which leaves the rest of the effective
Lagrangian invariant. Such a transformation however
introduces  the Yukawa term of the form
 \beq   \sqrt2 \{A_D + (\theta_{eff}/2\pi) A\}\,
\psi_M \psi_{\tilde M}. \label{nonlocal}\eeq 
 Since  the condensates $\bra A_D \ket
$ and $  \bra A \ket$
 are in general relatively complex, no phase rotation can eliminate the phases
completely from the Lagrangian: CP invariance is broken spontaneously in this case. 
It is interesting that   the above   shift of the dual
scalar $A_D$   transforms  Witten's boundary effect Eq.(\ref{elcharge}) - the
electric   charge of the magnetic monopole,  $\theta_{eff}/2\pi$,  - into the
standard  (albeit mutually non-local) minimal couplings of $M$  with $A_{D \mu}$ and
$A_{\mu}$, as can be  seen from the $N=2$ supersymmetric completion of the
Yukawa interaction, Eq.(\ref{nonlocal}).

One thus  reaches an amusing  conclusion that 
the massless theory depends (in a given vacuum) on the $\theta$ parameter, while the
massive theory is independent of  it!

\smallskip

The low energy CP invariance and non-renormalization of the $\theta$ parameter in the
massive case  may be   closely connected to the phenomenon of confinement.  According
to 't Hooft\cite{TH}, the confinement is a sort of dual superconductivity,  due to the
condensation of (color) magnetic charges.  Now if the dynamics of magnetic 
condensation is such that the magnetic monopoles  must have rigorously  zero electric
charge to be able to condense,  then it follows that  by    Witten's  formula 
Eq.(\ref{elcharge})   the {\it low energy}  $\theta$ parameter must be exactly   zero.
(For somewhat related ideas  see Ref.\cite{Schier}.)  
This seems to be precisely what happens in the massive Seiberg-Witten model. 
However, since  the independence of the low energy theory on the bare $\theta$
parameter is due to the presence of a massless fermion in the original model, an aspect
probably not shared by the ordinary QCD,   it is not clear whether  the massive
Seiberg -Witten model can be regarded as a good model of solution of the strong CP
puzzle.

Let us conclude with a general comment. 
The bare $\theta$ parameter (which can be set to zero) is renormalized in the infrared 
 by  multi-instanton effects 
(or loops of dyons in the dual variables)  differently in various vacua, i.e., 
 to a nonzero value in a generic
 vacuum  of the $N=2$ theory, to zero in the confining phase ($N=1$ theory).
Precisely these instanton effects are    responsible
for maintaining at any scale  the duality relation  $ \tau_D= - {1 / \tau},\,$  $\tau=
{\theta_{eff}/2\pi} + {4\pi i/g_{eff}^2},$  which generalizes the Dirac's quantization
condition\cite{Dirac} $ g e = 2 \pi n, \quad  n=0,1,2, \ldots. $ 
Note  how an old puzzle related to
the  Dirac's quantization condition 
({\it how  to maintain the quantization condition for $g$ and $e$ which are
both $U(1)$ coupling constants  hence which get renormalized smoothly in the same
direction as the scale is slowly varied?}) \cite{Zum} 
  is  solved in the Seiberg-Witten model. The "electric" coupling constant 
$g_{eff}$ here is truly a non Abelian charge and gets renormalized in the opposite
way compared to the $U_D(1)$  magnetic charge $g_D$. 
This consideration seems to strengthen  the idea  that magnetic monopoles and dyons
can appear in Nature  only as composite, solitonic particles  in the context of
a non Abelian gauge theory,   spontaneously broken  (or gauge-projected) to a group
involving  $U(1)$ subgroups.   

\smallskip

\noindent{\bf Acknowledgments}

One of the authors (K.K.)  thanks  LPTHE,  Centre d'Orsay, Universit\'e de
Paris-Sud where his study on these problems  started,  for a warm hospitality,  
the organizer and partecipants of the Kyoto
workshop on Supersymmetry (March 96) for an occasion to present and
discuss  some of the subjects  treated here, and  his collegues A. Di Giacomo, T.
Eguchi, Riccardo Guida, T. Kugo,  G. Paffuti, P. Rossi, Hiroshi Suzuki,  E. Tomboulis
and   Sung-Kil Yang for  
discussions.

\noindent{\bf Figure Caption}

\noindent{ Fig.1}    Mass spectrum of some stable  particles with magnetic charge
in the  $N=2$ theory as $u$ varies as
 $u=  e^{i  \alpha }\Lambda^2, \,\, \alpha = 0 \to
\pi.  $ The numbers near each curve indicate $(n_m, n_e)$. The unit of mass is 
$4 |\Lambda|/\pi.$

\begin{figure}[h]
\begin{center}
\leavevmode
\epsfxsize=10cm
\epsffile{kenplot.eps}
\end{center}
\caption{}
\end{figure}


\begin{thebibliography}{30}
\bibitem{SW} 
  N. Seiberg and E. Witten, Nucl. Phys. {\bf B426} (1994) 19;
\bibitem {TH} G. 't Hooft, Nucl. Phys. {\bf B190}[FS3] (1981) 455; 
\bibitem{Sei} N. Seiberg and K. Intrilligator, Nucl. Phys. {\bf B431} (1994)
551;
\bibitem{KK} K. Konishi, Phys. Lett. {\bf B135} (1984) 439; 
 \bibitem{RD}   R. F. Dashen, Phys. Rev {\bf 183} (1969) 1245;  {\it ibid} {\bf D3}
(1971) 1879; 
\bibitem{Ven} G. Veneziano, Phys. Lett.  {\bf 128B} (1983) 199 
\bibitem{Shifman} M.A. Shifman and A.I. Vainshtein, Nucl. Phys. {\bf B359}
(1991) 571;
 \bibitem{Ferrari} U. Lindstr\"om and M, Rocek, Phys. Lett. {\bf
355B} (1995) 492, A. Fayyazuddin,  Mod. Phys. Lett. {\bf A10} (1995) 2703, 
P. Argyres, A. Faraggi and A. Shapere, hep-th/9505190; M. Matone,  hep-th/9506181,
F. Ferrari and A. Bilal, hep-th/9602082; 
 \bibitem{Lee} T.D. Lee, Physics Reports C {\bf 9} (1974) 143;
\bibitem{Schier} G. Schierholz, Nucl. Phys. B(Proc. Suppl.)
 {\bf 37A} (1994) 203;
\bibitem{Dirac} P.A.M. Dirac,  Proc. Roy. Soc. {\bf A133} (1931) 60;
\bibitem{Zum} 
 B. Zumino, Erice Lectures (1966), Ed. A. Zichichi; 
S. Coleman,  Erice Lectures (1977), Ed. A. Zichichi.

\end{thebibliography}
\end{document}